\documentclass{IEEEtran}
\def\MJD{{\sc Majorana Demonstrator}}
\def\MJ{{\sc Majorana}}
\def\DEM{{\sc Demonstrator}}

\usepackage[switch]{lineno}
\usepackage[normalem]{ulem} 
\usepackage[dvipsnames]{xcolor}
\usepackage{cite}
\usepackage{amsmath,amssymb,amsfonts}
\usepackage{graphicx}
\usepackage{textcomp,nicefrac}
\usepackage{url}
\usepackage{hyperref}
\def\BibTeX{{\rm B\kern-.05em{\sc i\kern-.025em b}\kern-.08em
T\kern-.1667em\lower.7ex\hbox{E}\kern-.125emX}}
\begin{document}
\title{ADC Nonlinearity Correction for the {\sc Majorana Demonstrator}}
\author{N. Abgrall, J. M. Allmond, I. J. Arnquist, F. T. Avignone III au2, A. S. Barabash, C. J. Barton, F. E. Bertr, , B. Bos, M. Busch, M. Buuck, T. S. Caldwell, C. M. Campbell, Y-D. Chan, C. D. Christofferson, P. -H. Chu, M. L. Clark, H. L. Crawford, C. Cuesta, J. A. Detwiler, A. Drobizhev, D. W. Edwins, Yu. Efremenko, H. Ejiri, S. R. Elliott, T. Gilliss, G. K. Giovanetti, M. P. Green, J. Gruszko, I. S. Guinn, V. E. Guiseppe, C. R. Haufe, R. J. Hegedus, R. Henning, D. Hervas Aguilar, E. W. Hoppe, A. Hostiuc, M. F. Kidd, I. Kim, R. T. Kouzes, A. M. Lopez, J. M. Lopez-Castano, E. L. Martin, R. D. Martin, R. Massarczyk, S. J. Meijer, S. Mertens, J. Myslik, T. K. Oli, G. Othman, W. Pettus, A. W. P. Poon, D. C. Radford, J. Rager, A. L. Reine, K. Rielage, N. W. Ruof, M. J. Stortini, D. Tedeschi, R. L. Varner, S. Vasilyev, B. R. White, J. F. Wilkerson, C. Wiseman, W. Xu, C. -H. Yu, B. X. Zhu, and B. Shanks
\thanks{Manuscript received September 18, 2020; revised November 9, 2020; accepted November 19, 2020. Date of publication December 9, 2020; date of current version March 15, 2021. This work was supported in
part by the U.S. Department of Energy, Office of Science, Office of Nuclear Physics under Contract DE-AC02-05CH11231, Contract DE-AC05-00OR22725, Contract DE-AC05-76RL0130, Contract DE-FG02-97ER41020, Contract DE-FG02-97ER41033, Contract DE-FG02-97ER41041, Contract DE-SC0012612, Contract DE-SC0014445, Contract DE-SC0018060, Contract LANLE9BW, and Contract LANLEM77; in part by the Particle Astrophysics Program and Nuclear Physics Program of the National Science Foundation under Grant MRI-0923142, Grant PHY-1003399, Grant PHY1102292, Grant PHY-1206314, Grant PHY-1614611, Grant PHY-1812409, and Grant PHY-1812356; in part by the U.S. Department of Energy through the Lawrence Berkeley National Laboratory (LBNL)/Laboratory Directed Research and Development (LDRD), the Los Alamos National Laboratory (LANL)/LDRD, and the Pacific Northwest National Laboratory (PNNL)/LDRD Programs; in part by the Russian Foundation for Basic Research under Grant 15-02-02919; in part by the Natural Sciences and Engineering Research Council of Canada, under Grant SAPIN-2017-00023; in part by the Canada Foundation for Innovation John R. Evans Leaders Fund; in part by the Oak Ridge Leadership Computing Facility, Oak Ridge National Laboratory; in part by the National Energy Research Scientific Computing Center; and in part by the U.S. Department of Energy Office of Science User Facility.

Please see the Acknowledgment section of this  article for the author affiliations.}
}

\maketitle

\begin{abstract}
Imperfections in analog-to-digital conversion (ADC) cannot be ignored when signal digitization requirements demand both wide dynamic range and high resolution, as is the case for the \MJD~$^{76}$Ge neutrinoless double-beta decay search. Enabling the experiment’s high-resolution spectral analysis and efficient pulse shape discrimination required careful measurement and correction of ADC nonlinearities. A simple measurement protocol was developed that did not require sophisticated equipment or lengthy data-taking campaigns. A slope-dependent hysteresis was observed and characterized. A correction applied to digitized waveforms prior to signal processing reduced the differential and integral nonlinearities by an order of magnitude, eliminating these as dominant contributions to the systematic energy uncertainty at the double-beta
decay Q value

\end{abstract}


\section{Introduction}
\label{sec:intro}

\IEEEPARstart{S}{pectroscopic} measurements requiring signal digitization with both wide dynamic range and high energy resolution must pay special attention to nonlinearities in analog-to-digital conversion (ADC). The \MJD~\cite{Abgrall:2014} is such an experiment, consisting of an array of enriched high-purity germanium detectors (HPGe) used to search for the neutrinoless double-beta ($0\nu\beta\beta$) decay of $^{76}$Ge. This hypothetical nuclear decay emits two electrons without the balancing emission of anti-leptons; the observation of such a matter creation process would signify that lepton number is not conserved, with implications for the matter-antimatter asymmetry of the universe~\cite{Langacker:1988}. The \DEM\ requires a wide dynamic range for detection of low-energy spectral features of backgrounds for high-energy $0\nu\beta\beta$ decay signals. 
A wide dynamic range also enables searches for other Beyond-the-Standard-Model physics at low energy~\cite{Abgrall:2017,Alvis:2018}. High resolution is required for efficient pulse shape discrimination (PSD) of gamma and alpha radiation, and for careful measurement of the signal amplitude (energy) to distinguish $0\nu\beta\beta$ decay from the Standard Model process in which two neutrinos are emitted and from contaminating background events. Recently, the \MJ\ collaboration published its  $0\nu\beta\beta$ decay search results~\cite{Aalseth:2017, caldwell2018}, demonstrating high-efficiency PSD, very low background, and the best energy resolution to date among large-scale $0\nu\beta\beta$ decay searches. 
This achievement was made possible in part by the novel method presented in this paper for measuring and correcting ADC nonlinearities in the Gamma-Ray Energy Tracking In-beam Nuclear Array (GRETINA) Digitizer Modules~\cite{GRETINA,Anderson:2009} employed in the \MJD.

The experiment is staged at the 4850-foot level of the Sanford Underground Research Facility~\cite{Heise_2015} in Lead, SD. It is composed of 58 p-type point contact (PPC) high purity germanium detectors divided between two compact arrays housed within identical low-background cryostats. Each of the two detector arrays contains seven strings, with each string being an assembly of three, four, or five vertically stacked detectors. The PPC technology~\cite{Luke:1989, Barbeau:2007} was selected because of its superb energy resolution and ability to distinguish between multi- and single-site interactions~\cite{Alvis:2019dzt}. Charge collection in PPC detectors occurs on time scales of hundreds of nanoseconds to several microseconds~\cite{Mertens_2019}.

The point contact of the detector is connected by a spring-loaded pin to the gate of a field-effect transistor (FET) mounted on a low-mass front end (LMFE) board made of high-radiopurity materials~\cite{ABGRALL2015654}. In addition to the FET this circuit incorporates an amorphous-Ge feedback resistor, and the proximity of its traces provides the appropriate feedback capacitance for the charge-based amplification of the detector signals. The circuit also includes an additional capacitive-coupled trace for sending test pulses to the gate of the FET. The LMFE is located close to the detector in order to minimize stray input capacitance. The RC constant of the feedback loop is on the order of milliseconds.

The rest of the preamplifier lies outside the cryostat and is connected to the LMFE by a long (2.15~m) length of cable~\cite{ABGRALL2015654}. The voltage at the first stage of the preamplifier is measured at regular intervals to monitor temperature and leakage current stability. The second stage of the preamplifier is AC coupled to the first stage, and has two differential outputs which differ in gain by a factor of $\sim$3 (high gain and low gain). The detector signals have a sharp rising edge, the structure of which provides information on the charge drift, and a tail that falls exponentially with a $\sim$70~$\mu$s time constant arising from the AC coupling between the first and second stages of the preamplifier. 

For each crystal array, four circuit boards (``controller cards'') interface with the preamplifiers. Each of these controller cards contains sixteen 12-bit ADCs for monitoring baseline voltages and sixteen 16-bit digital-to-analog converters (DACs) for pulsing the FETs. The pulsers allow distribution of pulses of programmable amplitude and frequency to specified sets of FETs. These pulsers are used to monitor gain stability, trigger efficiency, and detector livetime. The pulsers can also be used for validation of digitizer linearity, as discussed below.

The high and low gain outputs of each detector's preamplifier are connected to  separate digitization channels. The GRETINA Digitizer Modules provide 10 channels per card, each with differential input and a 14-bit ADC digitizing at 100 MHz. The input dynamic range is $\pm$1.25~V. An on-board field-programmable gate array (FPGA) performs digital discrimination and trapezoidal shaping~\cite{Jordanov:1994}. The digitizers provide various triggering modes and accomplish raw data storage of triggered signals with a FIFO (first in, first out) memory. A mode in which a set number of samples are summed together during selected portions of the trace before writing to memory allows extension of the captured time window at a lower sampling period while staying within the FIFO's 4~kB event record size~\cite{Alvis:2018}. The digitization electronics for the two cryostats operate in separate Versa Module Eurocard (VME) crates, each housing the requisite number of GRETINA digitizer boards and a single board computer (SBC) to read out the digitizers in that crate. The two SBCs communicate with one central computer running ORCA (Object Oriented Real-time Control and Acquisition) which controls the entire data acquisition (DAQ) system~\cite{Howe:2004}. All acquisition parameters are programmable and easily accessed through an ORCA interface.

Full analysis of the pulse shapes is performed offline. The nonlinearity correction is applied to digitized waveforms prior to other signal processing routines. Signal amplitudes are measured to extract event energies using a trapezoidal filter~\cite{Jordanov:1994,Johns:1997} with a $4~\mu$s integration time, a flat-top of $2.5~\mu$s, and employing a modified pole-zero adjustment to correct for charge trapping~\cite{Alvis:2019dzt, ChargeTrapping:2019}. The smoothed derivative of the pulse is computed with a current estimator based on the slope of a running linear fit to a 100-ns range of the waveform. This estimator is used to distinguish signal-like single-site events, which show a sharp single peak in the current, from multi-site background interactions, dominated by multiple scattering of gammas, which exhibit multiple current peaks and/or a suppressed maximum current. A third pulse shape parameter, the ``delayed charge recovery'' (DCR), looks for an anomalous slope in the exponential tail of the pulse, and is used to discriminate against surface alpha background interactions~\cite{Alvis:2018,Arnquist:2020veq}. The slope of the tail is calculated as the difference of two 1~$\mu$s integration regions, one starting 2~$\mu$s after the waveform reaches 97\% of its maximum and one at the end of the waveform, divided by the intervening time interval. Alpha interactions result in extra charge collected during this delayed interval, and are identified as pulses having an anomalous tail slope in their raw signal. Our application of these pulse shape analysis (PSA) algorithms rely on highly linear analog-to-digital conversion of the detector signals, as is quantitatively shown in Sections IV and V.

Periodic nonlinearities have been observed in the ADC chips used in the GRETINA Digitizer Module (Analog Devices AD6645) arising from the subranging nature of the ADC implementation. Of particular note is that the nonlinearities in these ADCs depend not only on the voltage level but also the rate at which the voltage changes~\cite{Dallet:2005}. Uncorrected, these ADC nonlinearities affect energy determination by up to several keV, on the same order as the 2.5~keV full width at half maximum (FWHM) of the $0\nu\beta\beta$ decay peak. This would require the search region for the $0\nu\beta\beta$ decay peak to be made much larger than would otherwise be necessary, increasing the background.  For the pulse shape discrimination parameters, nonlinearities result in energy dependence of the signal acceptance and background rejection that increase uncertainty and complicate spectral analysis. In the case of DCR, the fluctuations with energy are on the same order as the ~1\% signal sacrifice, making ADC nonlinearity a dominant contributor to efficiency uncertainty prior to correction.

A number of methods are available for correcting nonlinearity, for example the histogram method~\cite{Linnenbrink:2001}, integral nonlinearity curve tables~\cite{Suchanek:2008}, using the analytic inverse of the integral nonlinearity curve~\cite{Suchanek:2009}, and the blind calibration algorithm~\cite{Gande:2014}. However, most methods require special equipment or architectures, and/or lengthy measurement campaigns. They also often assume that ADC differential nonlinearities are fixed constants that are independent of the time variation of the input signal. These aspects made standard methods inadequate for the \MJD.

In this paper, we present the nonlinearity correction developed for the \MJD. First we will describe the measurement of the nonlinearities by applying external signals and measuring the response of each digitizer channel. Then we describe our nonlinearity correction algorithm, and quantify the energy performance of the \DEM\ after applying the nonlinearity correction. Finally, we discuss the validation of the linearity of the corrected energy spectrum using external pulsers. 

\section{Nonlinearity Measurement}
\label{sec:measurement}
A measurement of the differential nonlinearity (DNL) of an ADC is often accomplished by applying a precise voltage ramp that spans the entire input range of the ADC, and counting how often each ADC code appears in the resulting output stream. For each output code, the fractional deviation from the mean frequency yields the differential nonlinearity, and integration of the DNL yields the integral nonlinearity (INL)~\cite{Johns:1997}. This procedure requires the use of a highly linear, precise signal generator, since any nonlinearities in the voltage ramp will be transferred as an apparent DNL in the measurement.  No such precision generator was available in the underground laboratory where we performed our measurements, so we devised a modified procedure where our measurement of ADC nonlinearity is performed through application of two separate ramped voltage signals to the differential inputs of a digitizer channel. As described in detail below, this allows a precise determination of the DNL with the use of two low-cost signal generators. In a way similar to the standard method described above, we use the fractional deviation of the observed occurrence of each ADC code from its expected occurrence to define the differential nonlinearity, and the cumulative sum of the resulting DNL to define the integral nonlinearity~\cite{Johns:1997}.

\begin{figure}[htbp]
\centering
\includegraphics[width=0.47\textwidth]{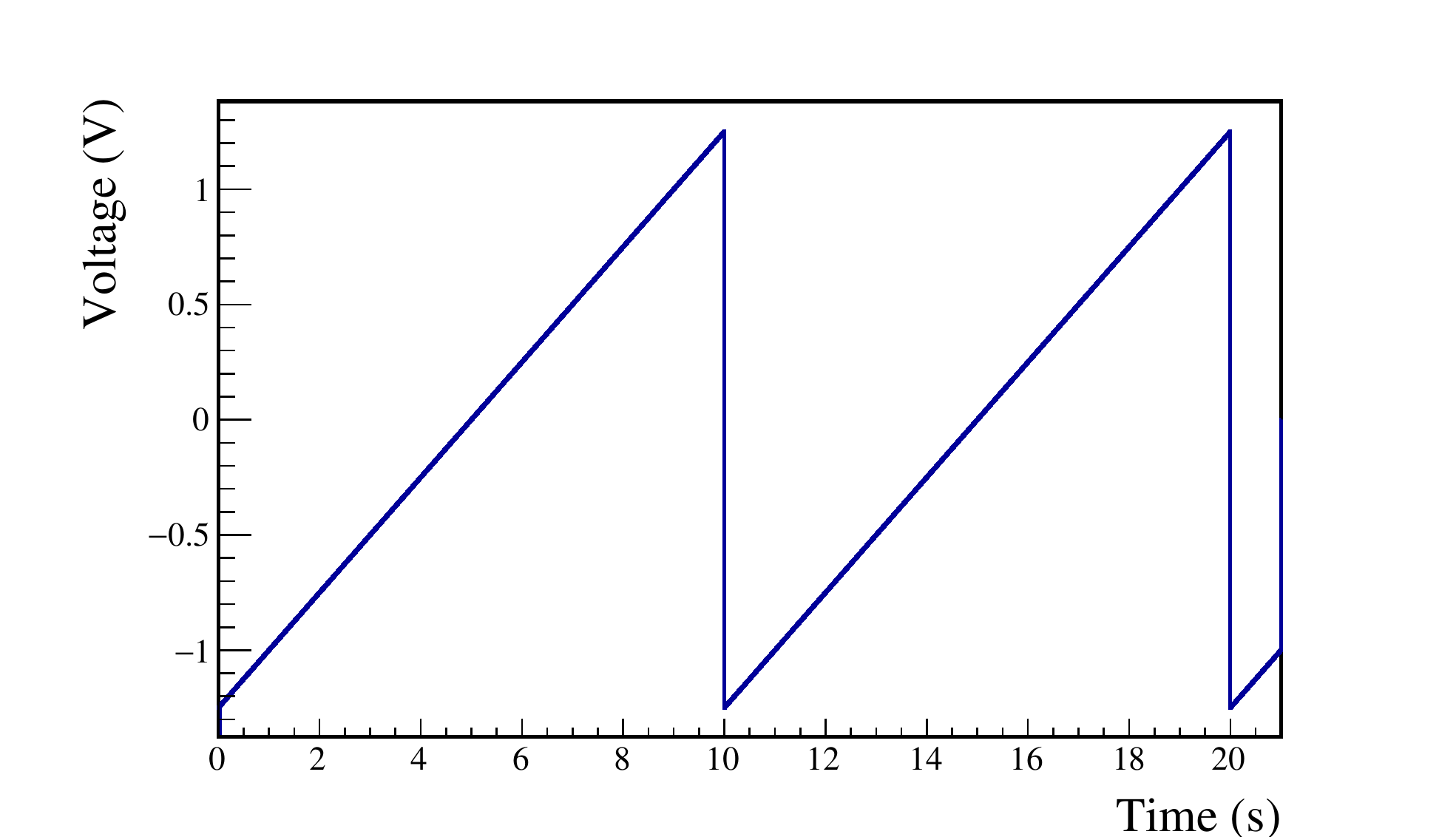}
\includegraphics[width=0.47\textwidth]{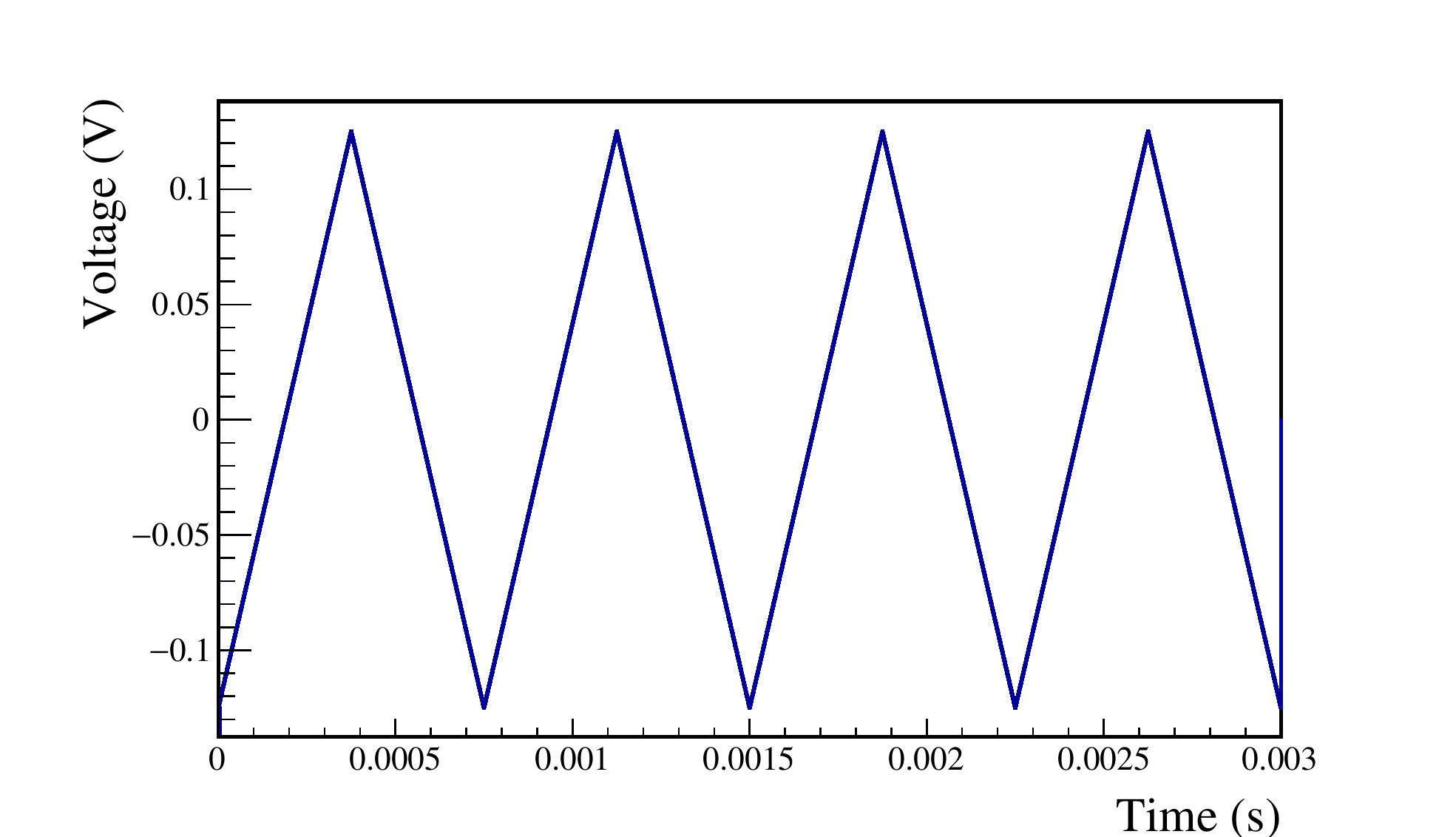}
\caption{Schematic plot for the external input: Top) slow ramp signals and Bottom) fast ramp. 
}
\label{fig:Schematic_Input} 
\end{figure}

The measurement procedure makes use of two external function generators (Agilent 33220A) applied to the two differential inputs of a digitizer channel. The first function generator provides a slow ramp covering the full ADC range, while the second provides a faster ramp of smaller amplitude that effectively modulates the signal from the slower ramp. In our setup, the slow ramp waveform is set to 100\% symmetry (generating a sawtooth wave) with a period of 10~s and an amplitude of $\pm$1.25~V. The fast ramp waveform is set at 50\% symmetry (triangle wave) with a period of 750~$\mu$s and an amplitude of $\pm$125~mV. Figure~\ref{fig:Schematic_Input} shows a schematic plot of the two function generator outputs. These signals are summed together by the differential inputs of the digitizer. 

\begin{figure}[htbp]
\centering
\includegraphics[width=0.45\textwidth]{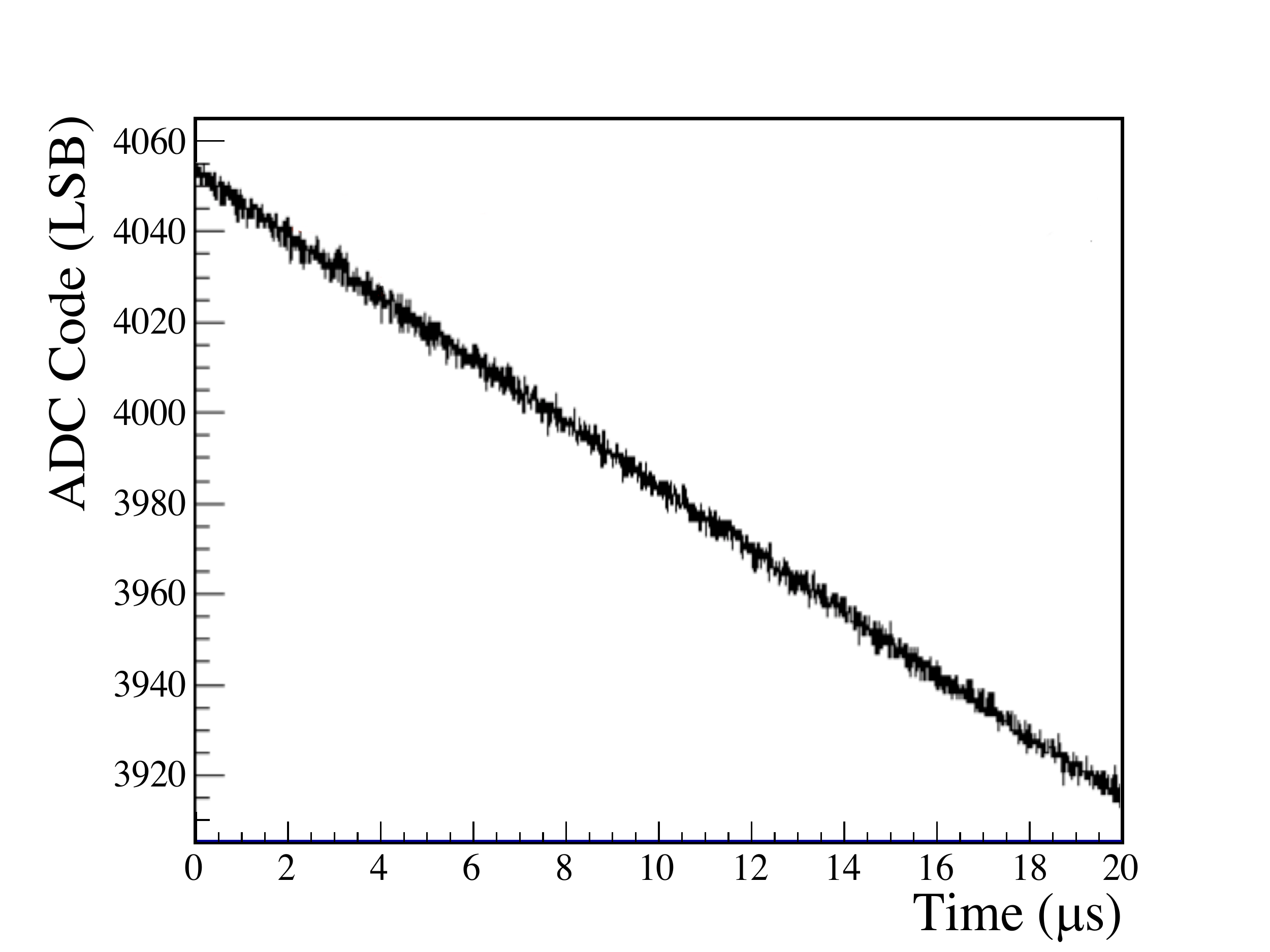}
\caption{A typical measured (down-going) waveform.}
\label{fig:NL_WF} 
\end{figure}
A synchronized output of the faster function generator was used to externally trigger the digitizer card on a rising or falling portion of the fast ramp. The digitized traces record a short, monotonic 20-$\mu$s ($sim$2000-sample) region of the fast ramp, as shown for example in Fig.~\ref{fig:NL_WF}. The slope of the waveform is determined by the slope of the fast ramp, and its overall ADC offset is determined by the location along the slow ramp. In the example of Fig.~\ref{fig:NL_WF}, each waveform spans about 130 ADC codes.

For a perfect ADC and linear voltage ramp, with no DNL or noise, a count of the ADC codes within any one waveform would produce a flat histogram bounded by the initial and final values corresponding to the net ADC input voltage. Each code within those bounds would appear $ 2000/130 \approx 15 $ times. In reality, the code counts are modulated by the ADC bin widths (i.e. the DNL) and the bounds of the histogram are rounded by noise. To remove the latter effect, we discard the code counts for ADC values close to the boundaries. To increase statistics and integrate out any nonlinearities in the fast ramp, we collect many such waveform segments and keep track of how many times each ADC code falls within the accepted range of any segment.

\begin{figure}[htbp]
\centering
\includegraphics[width=0.45\textwidth]{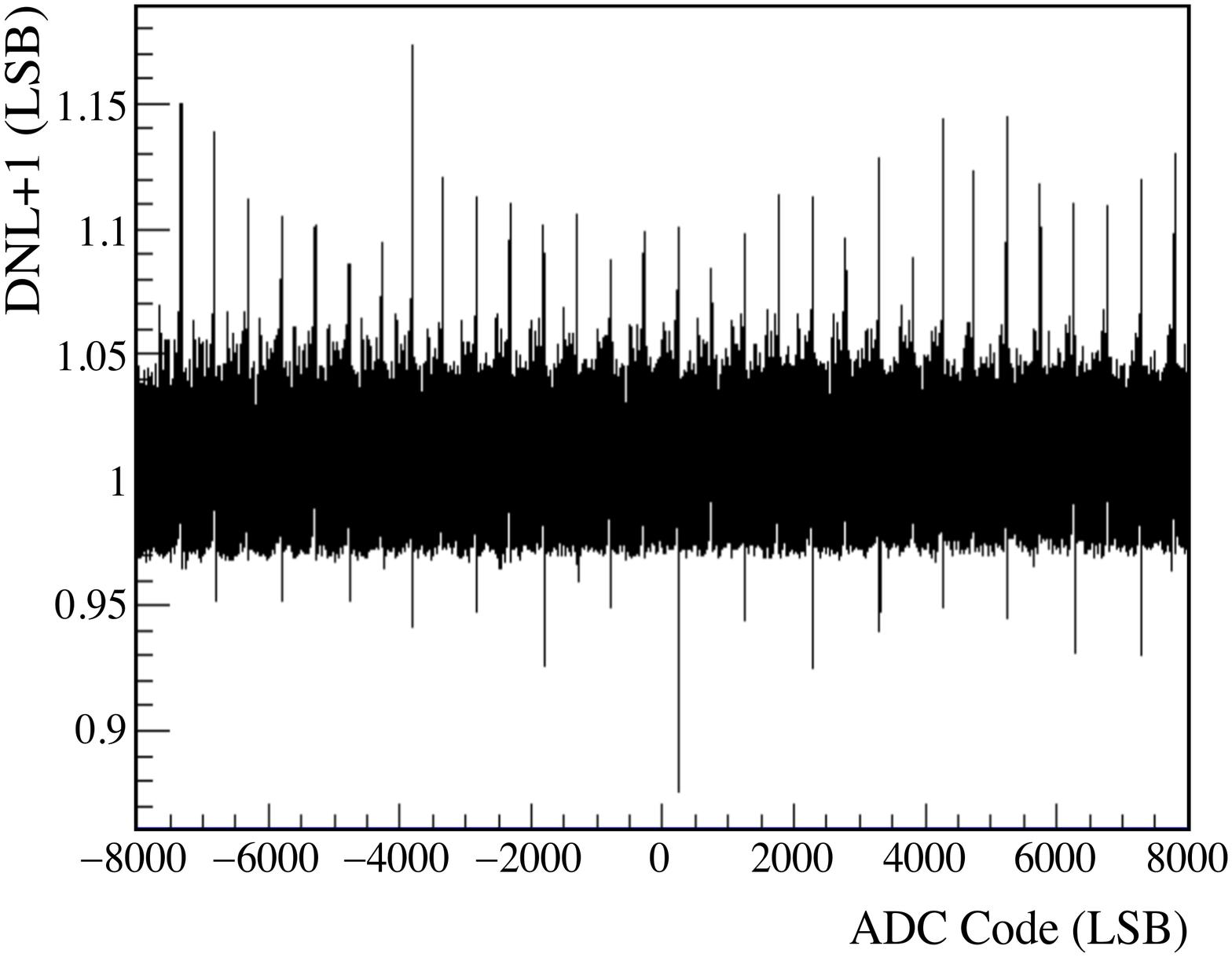}
\includegraphics[width=0.4\textwidth]{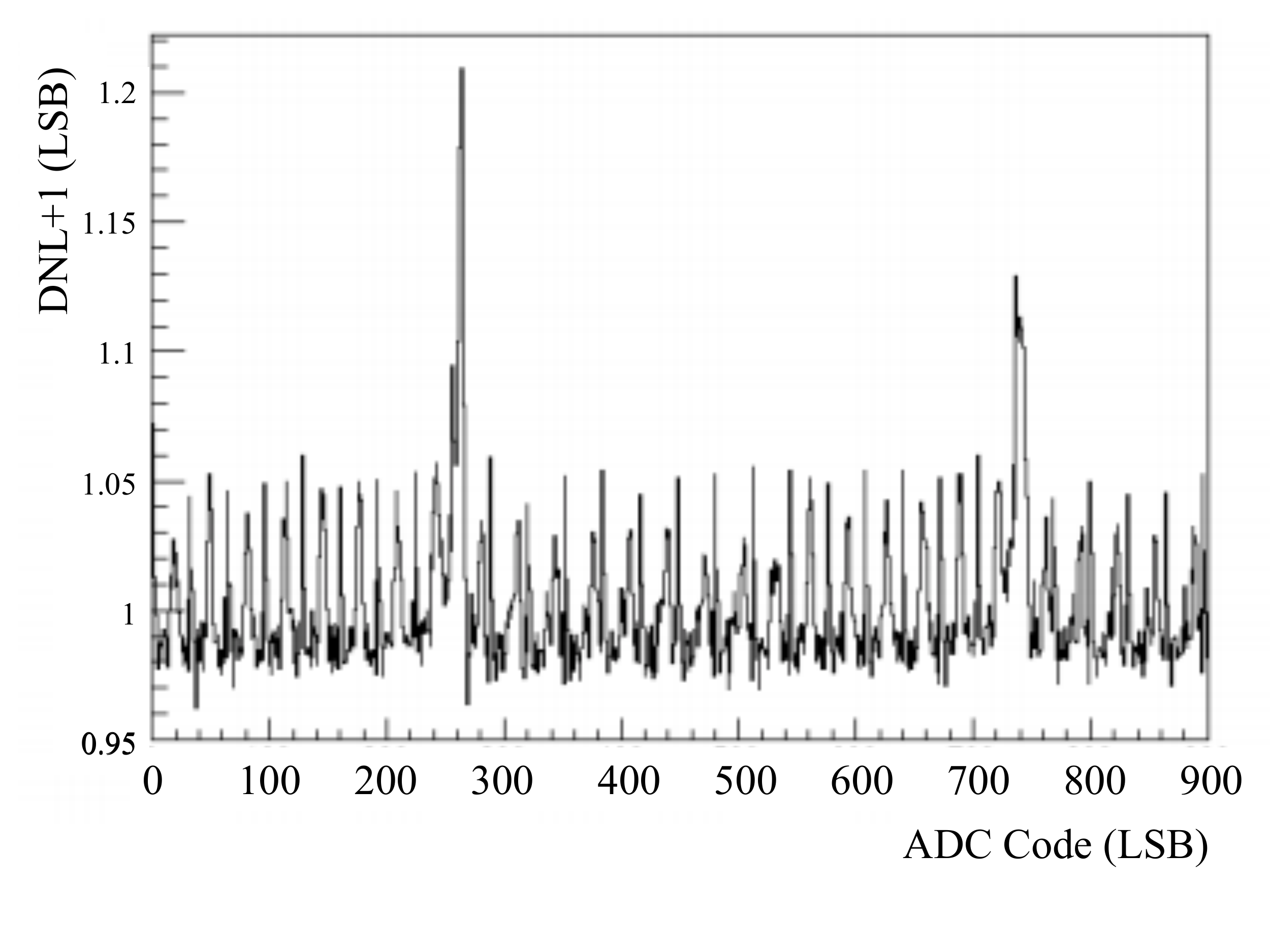}
\caption{(Top) An example measured average DNL curve at each ADC code using high statistics data of a single digitizer channel. (Bottom) A zoom-in of the DNL curve. The larger positive spikes have a height less than 1 LSB but a finite width, and result in the large INL steps that are greater than 1 LSB in height as shown in Fig.~\ref{fig:Measured_INL}.}
\label{fig:Measured_DNL}
\end{figure}

Data was taken for about 30 minutes, corresponding to roughly 2.4M waveforms recorded on each digitizer channel. For each ADC code, this yields an average of about 25000 segments where the code falls within the accepted range, and an average expected total count of about 0.4M. An example of the ratio of code frequency to the expected value is shown in Fig.~\ref{fig:Measured_DNL}, and corresponds to the DNL plus unity. The statistical uncertainty in each ratio is $ \sim 0.0016$.

As the slow ramp shifts the voltage offset of the fast ramp, each ADC code traverses varying locations along the fast ramp. As a result, the DNL we compute averages over any nonlinearities of the fast ramp, making our method insensitive to imperfections in the fast ramp linearity. The change in the slow ramp is much less than 1 ADC code (LSB) over the digitization window, and the computed DNL corrects for the measured frequency with which a particular ADC code is traversed, so that our method is also insensitive to any nonlinearities in the slow ramp. These features enable the use of lower-cost function generators without stringent linearity specifications. The method does however require a digitizer with differential inputs, and moderately time-consuming data acquisition.

\begin{figure}[htbp]
\centering
\includegraphics[width=0.48\textwidth]{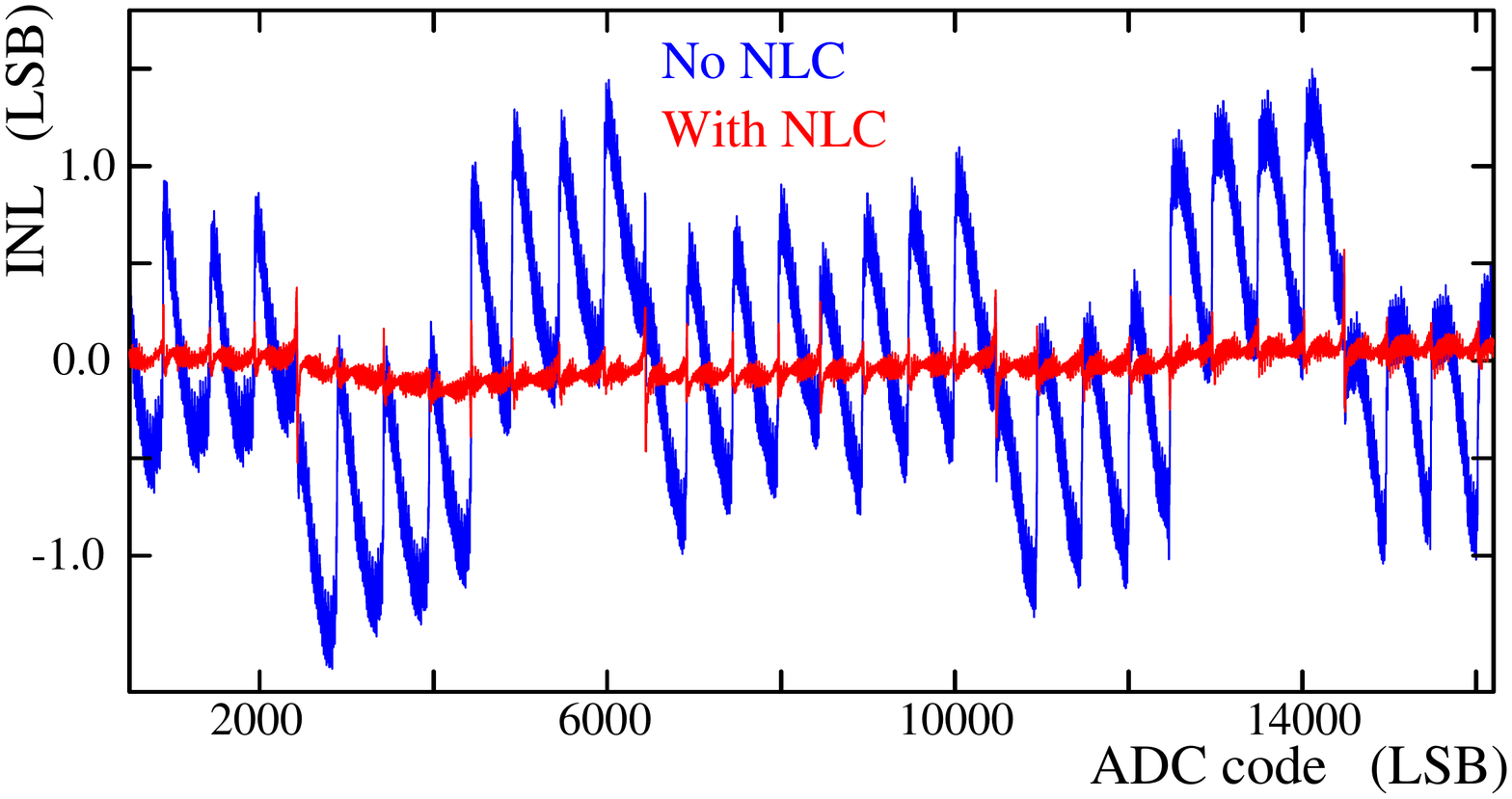}
\includegraphics[width=0.4\textwidth]{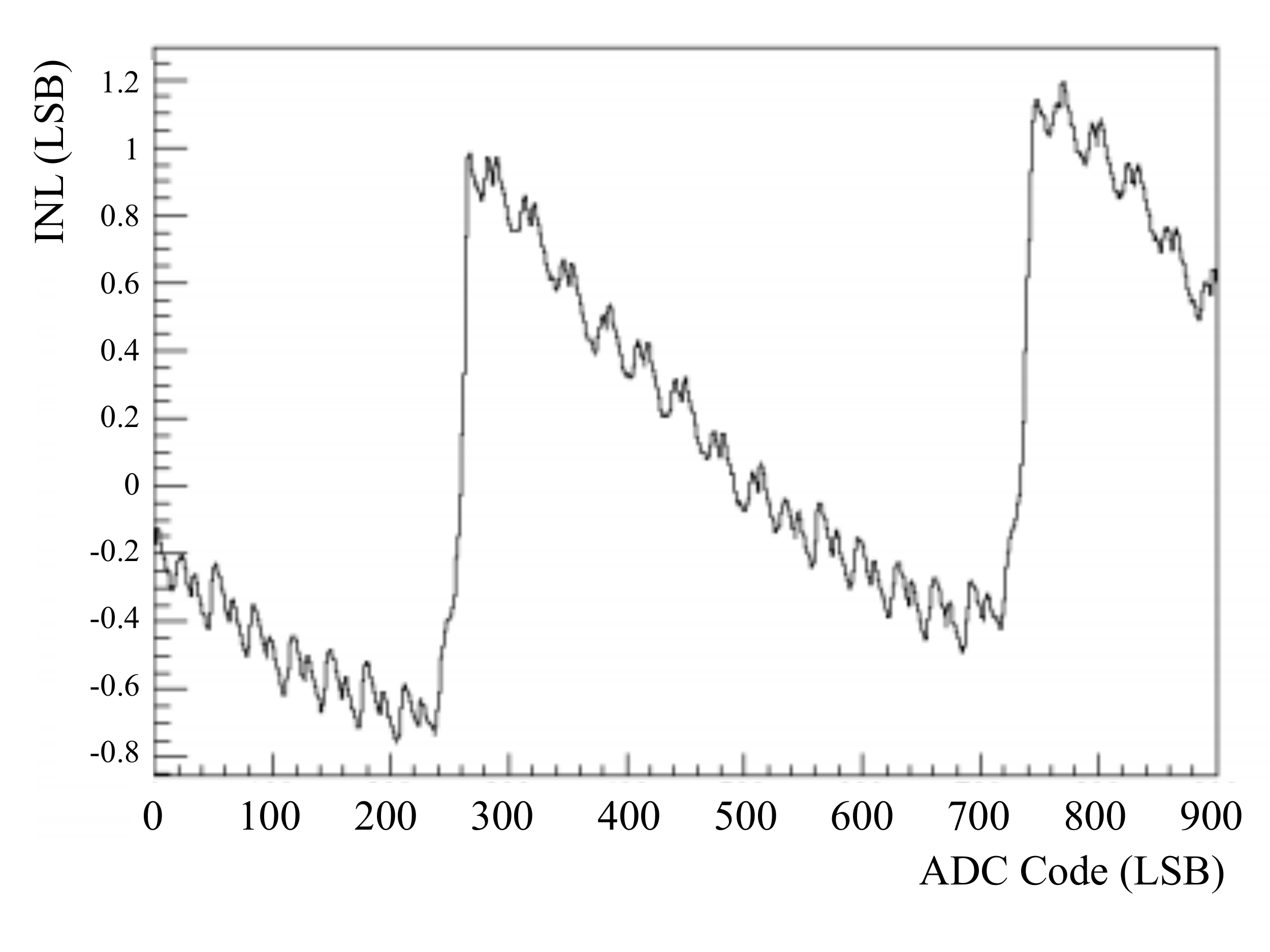}
\caption{(Top) An example measured INL curve for a single digitizer channel. The zig-zag pattern arises from the picket fence structure of the DNL curve. Blue dots show the INL before the nonlinearity correction while the red dots show the INL after our correction is applied. (Bottom) A zoom-in of the INL curve before the nonlinearity correction.}
\label{fig:Measured_INL}
\end{figure}

The measured DNL curve exhibits a picket-fence-like structure that is typical of multi-range ADCs. All digitizer channels measured show a similar trend with a similar magnitude; however the detailed structure and the sizes of the DNL spikes varied somewhat from one digitizer channel to the next, requiring their individual measurement. For each digitizer channel, we integrate the measured DNL curve, and fit and subtract away any overall slope to obtain the INL curve, as shown for example in Fig.~\ref{fig:Measured_INL}. The maximum deviation of the INL is at the level of approximately 2~ADC codes, corresponding to about 1~keV (3~keV) for high gain (low gain) detector signals in the \MJD. 

\begin{figure}[htbp]
\centering
\includegraphics[width=0.45\textwidth]{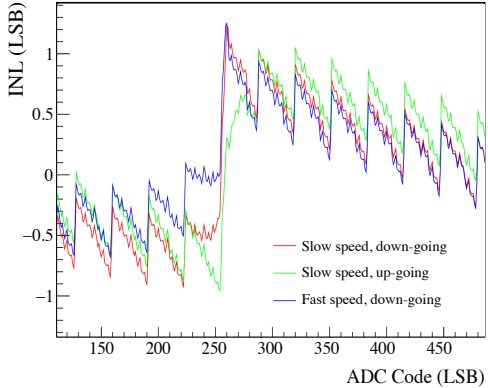}
\caption{
An example INL curve over a small ADC range for different fast ramp amplitudes and directions. The ADC exhibits a hysteresis effect, which can be seen in both the measured INL curve deviation between opposite ramp directions (shown by the green and red curves) and in the larger deviation seen at higher ramp rates (seen in the difference between the red and blue curves at ADC codes less than 260). The delayed response time constant describing the hysteresis effect is found from the number of ADC codes needed for the curves to come into agreement following a large nonlinearity. The ramp rate is used to convert this value into a time.
}
\label{fig:Measured_NL_zoom}
\end{figure}

\begin{figure}[htbp]
\centering
\includegraphics[width=0.4\textwidth,angle=-90]{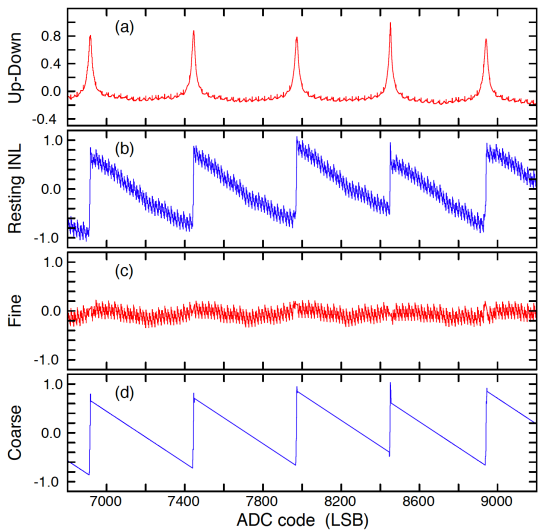}
\caption{From top to bottom: a) up-going-ramp minus down-going-ramp, b) the combined ``resting'' INL, c) the fine structure, and d) the coarse structure of the large INL steps joined with linear
segments. The hysteresis correction is applied only to this coarse structure, as described in the text.
}
\label{fig:diff_comb_part1_part2}
\end{figure}

Using this technique, we  measured the INL with varying amplitude and trigger positions (up-going vs.\ down-going regions) of the fast ramp. The resulting ADC INLs exhibit hysteresis, particularly in the vicinity of the large DNL peaks (see Fig.~\ref{fig:Measured_NL_zoom}). We interpret the hysteresis as a delayed response of the ADC to the signal. As shown in Fig.~\ref{fig:diff_comb_part1_part2}, the locations of the large INL steps can be determined by using the difference between the up-going ramps and the down-going ramps. After traversing the large INL step, the INL exponentially recovers to the ``resting'' value measured when large INL deviations are not traversed. That is, in Fig.~\ref{fig:Measured_NL_zoom} the resting INL is best estimated by the up-going ramps below the step, and by the down-going ramp above the step. For each trace, after the step the INL values can be seen to exponentially return to the resting values at a rate that depends on the ramp speed.

The origin of this time-delay effect involves the overall digitizer channel; however the ADC chip plays an important role since it is apparent only at the large steps in the INL. Using the slope of the fast ramp and the distance in ADC code for the INL curves to come into agreement with each other following a large nonlinearity, we roughly estimated the time scale of the delayed response to be $\sim$1.4~$\mu$s. This time delay is incorporated into the nonlinearity correction described below. Our method was sufficient to reduce ADC nonlinearities to a negligible level, however the time delay could be optimized further with more careful study.

\section{Nonlinearity Correction}

We now describe how these measured nonlinearity values are used to correct the detector waveforms (and thereby the event energies and PSA parameters) in the analysis of data taken in the experiment.

After searching for digitization problems such as range saturation, recorded waveforms are immediately corrected for nonlinearities prior to further digital signal processing. To first order, our correction is applied using the INL curve as a look-up table. Additionally, a small recursive adjustment is made to correct the observed time delay in the ADC response. 

First, we prepared estimates of the resting INL curves in which the distortion due to the time delay was removed. They are computed from INLs measured with both up-going ramps and down-going ramps. The difference between the two is used to locate the ADC codes where the major steps occur.
For each $\sim$500-code region between the major steps, the resting INL for the first 150 ADC codes is taken from the down-going ramp, the up-going ramp is used for the last 150 ADC codes, and a linear combination is used for the codes in between.  Linear fits to the range of values between the large steps then decompose the resting INL values into two subsets: a ``coarse'' set (${\rm INL}^{\rm c}$) of linear segments separated by the large steps, and a ``fine'' set (${\rm INL}^{\rm f}$) of the remaining fine structure without the large steps.

Once the resting INL curves are estimated, the nonlinearity correction is then applied. In the absence of the delayed response exemplified in Fig.~\ref{fig:Measured_NL_zoom}, the correction would be applied by simply subtracting the INL from each ADC code of the waveform. However, to account for the time delay effects during and following the rise of the detector signal, we need to keep track of the signal history. We first assume that the initial value of the detector signal (the “baseline”) is essentially static, so that there are no signal changes in the few $\mu$s prior to the triggered waveform that need to be accounted for.  This allows us to initialize the correction at the start of the detector waveform, and then keep track of time delays in the INL created as the signal rises and then falls.

We therefore compute
\begin{equation}
{\rm ADC}'_n = {\rm ADC}_n - {\rm INL}^{\rm c}_{{\rm ref},n} - {\rm INL}^{\rm f}_n,
\end{equation}
where ${\rm ADC}'_n$ is the corrected ADC code of waveform sample $n$, ${\rm ADC}_n$ is its original ADC code, and ${\rm INL}^{\rm c}_{{\rm ref},n}$ is a recursively computed reference correction that exponentially approaches the ADC code's coarse resting INL value with time constant $\tau$. It is given by
\begin{equation}
    {\rm INL}^{\rm c}_{{\rm ref}, n} = {\rm INL}^{\rm c}_{{\rm ref}, n-1} + ( {\rm INL}^{\rm c}_n - {\rm INL}^{\rm c}_{{\rm ref}, n-1} ) \frac{\tau_s}{\tau},
\label{eq:inlref}
\end{equation}
where $\tau_s$ is the sampling period (10 ns, $\tau_s \ll \tau$). The initial value of INL$_{{\rm ref},n}$ (INL$_{{\rm ref},0}$) is set to the INL value of the ADC code corresponding to the waveform baseline, taken as the average ADC code of the first 1~$\mu$s of the digitized signal. The value of $\tau$ is taken to be 1.4~$\mu$s for all digitizer channels; attempts to optimize this parameter code-by-code or channel-by-channel did not yield significant improvement in residual nonlinearities.

\section{Performance of the Nonlinearity Correction}
\label{sec:NLPerf}

\begin{figure}[htbp]
\centering
\includegraphics[width=0.45\textwidth]{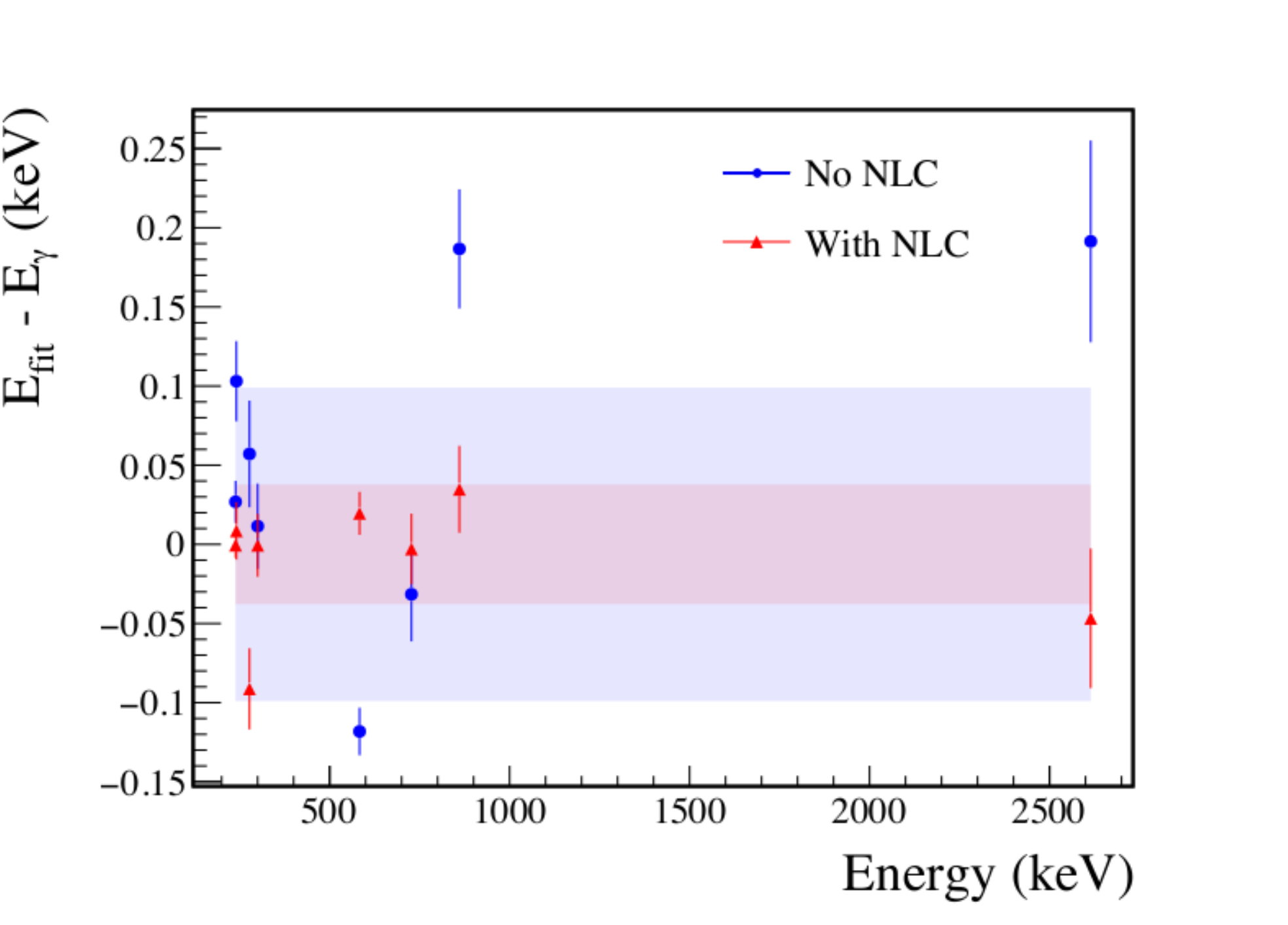}
\caption{An example of calibrated energy residuals in the high-gain amplification output of one detector before (blue circles) and after (red triangles) the nonlinearity correction. The shaded regions characterize the non-statistical spread of the data points.
}
\label{fig:NL_calibration}
\end{figure}

The most straightforward and direct method to measure the performance of the nonlinearity correction is to compare energy calibration residuals before and after the correction is applied. Figure~\ref{fig:NL_calibration} shows such a comparison for $^{228}$Th source data taken in high-gain channels with the \MJD\ calibration system~\cite{Calibration:2017}. The nonlinearity correction reduces the spread of the residues from 0.1~keV to 0.04~keV. Since other sources of energy uncertainty are on the order of 0.1~keV\cite{caldwell2018}, this comparison shows that without correction, ADC nonlinearity would be a significant source of energy uncertainty in the \MJD.

\begin{figure}[htbp]
\centering
\includegraphics[width=0.45\textwidth]{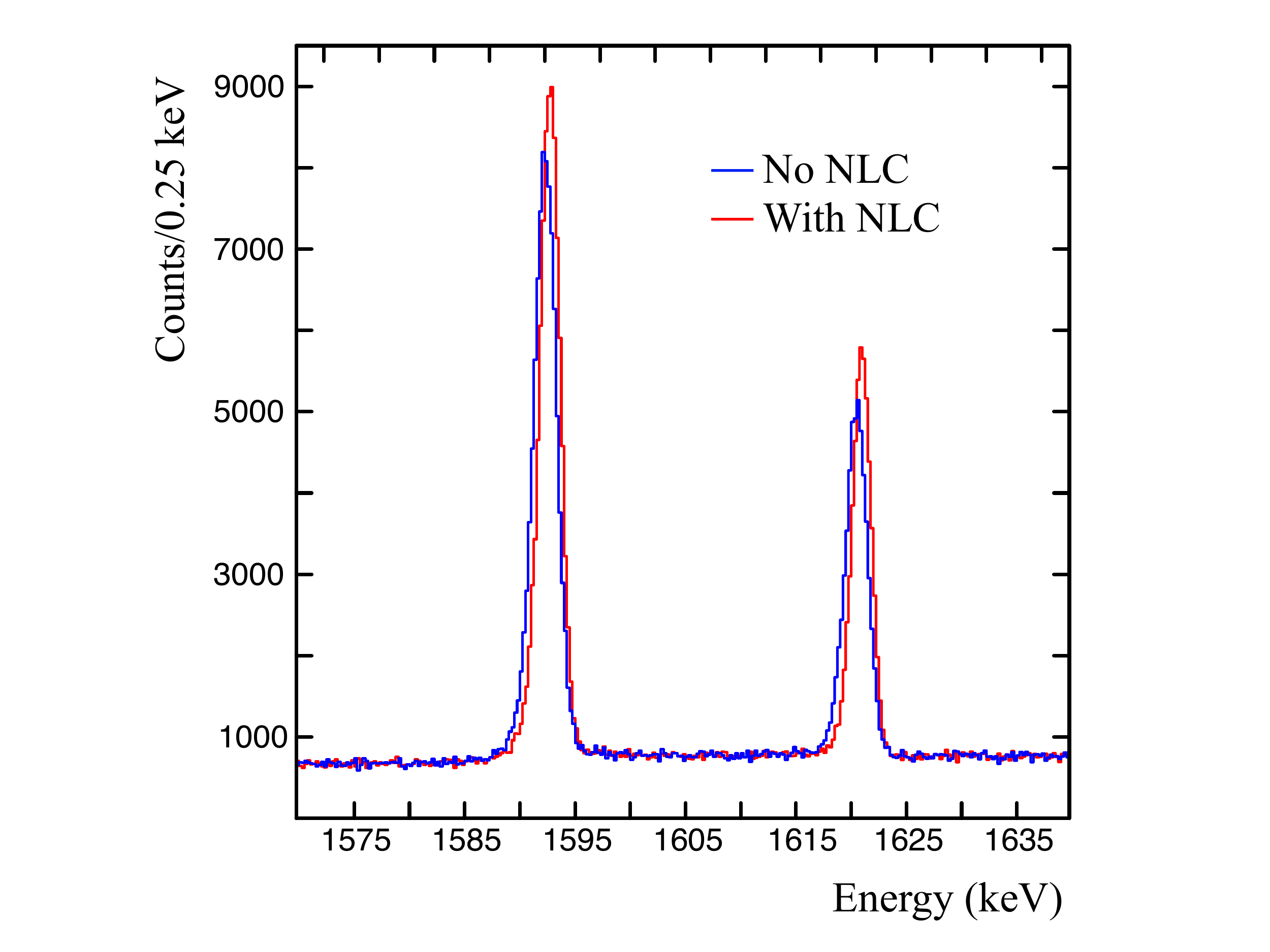}
\caption{
The 1592.5~keV DEP of the 2614.5-keV gamma emission of $^{208}$Tl and the 1620.7-keV $^{212}$Bi gamma ray observed in low-gain channels for $^{228}$Th calibration source data, with (red) and without (blue) the nonlinearity correction applied.}
\label{fig:peakdep}
\end{figure}

To give a demonstration of the improvement that is visible by-eye, Fig.~\ref{fig:peakdep} shows a zoom-in of the calibration spectrum for low-gain channels near the 1592.5 keV double escape peak (DEP) of the 2614.5-keV $^{208}$Tl gamma emission. This peak is not included in the calibration fits, and its single-site topology serves as a good proxy for the neutrinoless double-beta decay peak~\cite{Alvis:2019dzt}. The 1620.7-keV $^{212}$Bi gamma peak is also visible nearby. As can be seen in the figure, the nonlinearity correction improves both the location and width of the peaks by an even larger factor than for the high-gain channels: without the nonlinearity correction, the calibrated energy of the DEP is shifted down by 0.45~keV, and its FWHM is increased from 2.16~keV to 2.41~keV.

While Fig.~\ref{fig:NL_calibration} shows clearly that the nonlinearity correction improves the energy uncertainty, it does not provide a measure of the remnant contribution to the energy uncertainty of the ADC nonlinearity after applying our correction, because ADC nonlinearity is not the only source of variance in the data. Thus other methods are required to measure the performance of the correction in more detail.

\begin{figure}[htbp]
\centering
\includegraphics[width=0.45\textwidth]{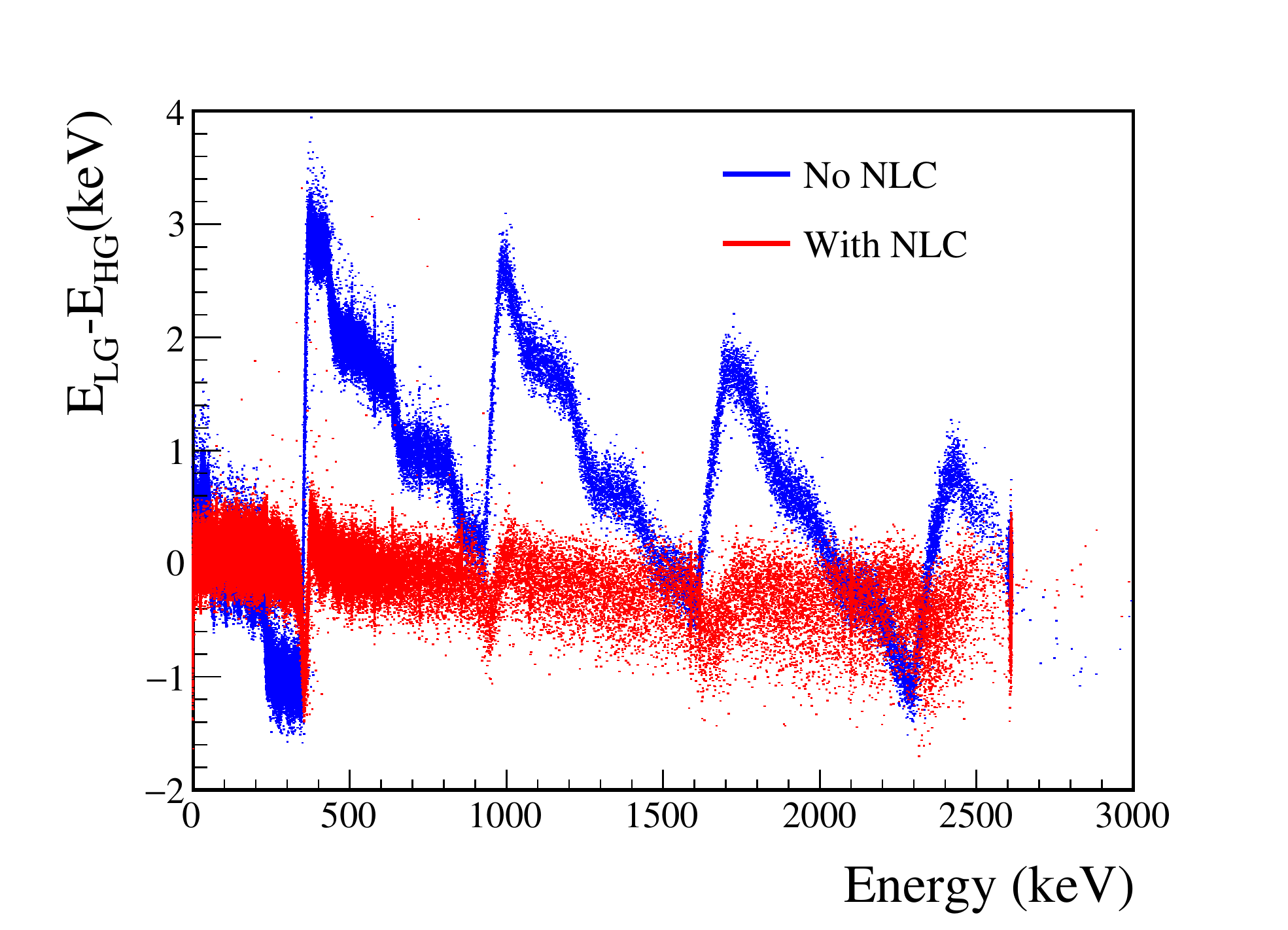}
\caption{The energy difference between the high gain and the low gain channels of $^{228}$Th calibration source events for one detector. Blue dots show the energy difference prior to nonlinearity correction while red dots show the energy difference after our correction is applied.}
\label{fig:NL_HGLG}
\end{figure}
Since low gain channels in the \MJD\ have $\sim$1/3 the gain of high gain channels, the nonlinearity has different amplitude and energy periodicity in the high and low gain channels. This allows one to use the energy difference between the low and high gains to investigate the nonlinearity. Figure~\ref{fig:NL_HGLG} shows one example of the difference in energies recorded simultaneously by each gain for events collected during a $^{228}$Th calibration source deployment. The uncorrected trend (blue) exhibits a superposition of sawtooth-shaped nonlinearities from the low gain (large amplitude, long period sawtooth) and high gain (small amplitude, short period sawtooth) channels.  Both patterns begin to wash out at higher energies because the signal region in the tail that is integrated by the trapezoidal filter spans a broader range of ADC codes,  averaging away the nonlinearity. After correction (red trend), the energy difference is reduced by roughly an order of magnitude. The features in the remnant nonlinearity, including the overall slope and the appearance of small structures, vary subtly from channel to channel. However, based on the relative gains, on average the low gain contribution to the remnant nonlinearity should be roughly three times that from the high gain channel. The observed patterns are consistent with this estimate.

\begin{figure}[htbp]
\centering
\includegraphics[width=0.45\textwidth]{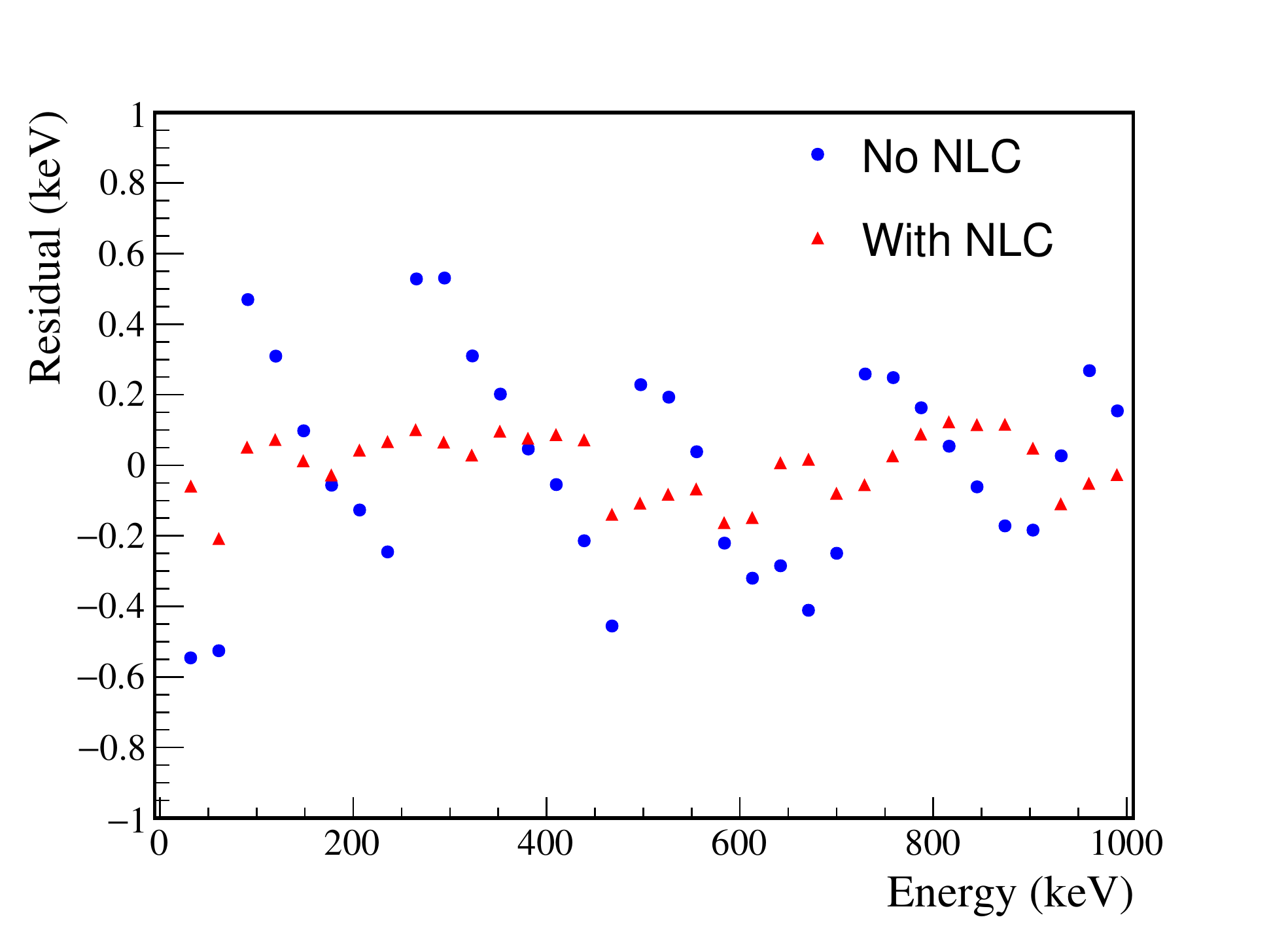}
\caption{Digitizer linearity as assessed via front-end pulser scans, with and without estimated energies corrected by the primary nonlinearity correction. The image displays residuals from a quadratic fit to the estimated energies of a sweep over evenly-spaced pulser amplitudes. Statistical uncertainties are smaller than the marker size.}
\label{fig:NL_FrontEndPulser}
\end{figure}

The front-end pulsers described in the introduction were also used to assess the ADC linearities over a limited dynamic range. A series of runs was taken in which the front-end pulser amplitude was stepped evenly from 0-1~MeV, and the amplitudes of the output pulses were measured with and without the nonlinearity correction applied. Observed nonlinearities cannot be distinguished as arising from the pulser or the digitizer itself. However, since the energy calibration function is a second order polynomial, nonlinearities up to second order in energy are calibrated away. After fitting and subtracting a second order polynomial trend, the uncorrected data exhibits a strong sawtooth trend indicative of the ADC nonlinearity, as shown for an example channel in Fig.~\ref{fig:NL_FrontEndPulser}. After applying the nonlinearity correction, the residuals are greatly reduced. To validate this method, the measurement was repeated with the pulsers attenuated by factors of 2, 5, and 10, so that nonlinearities inherent to the pulser itself would be attenuated by the same factor. The attenuated results matched the unattenuated measurements to within the typical point-to-point scatter visible in Fig.~\ref{fig:NL_FrontEndPulser}. From these sweeps, we conclude that below $\sim$1~MeV, remaining deviations from linearity are on the order of $\pm$0.1-0.3~keV in all detectors in both cryostats, and are consistent with the more conservative method based on energy differences between low and high gain data that extends to much higher energies.

\begin{figure}[htbp]
\centering
\includegraphics[width=0.45\textwidth]{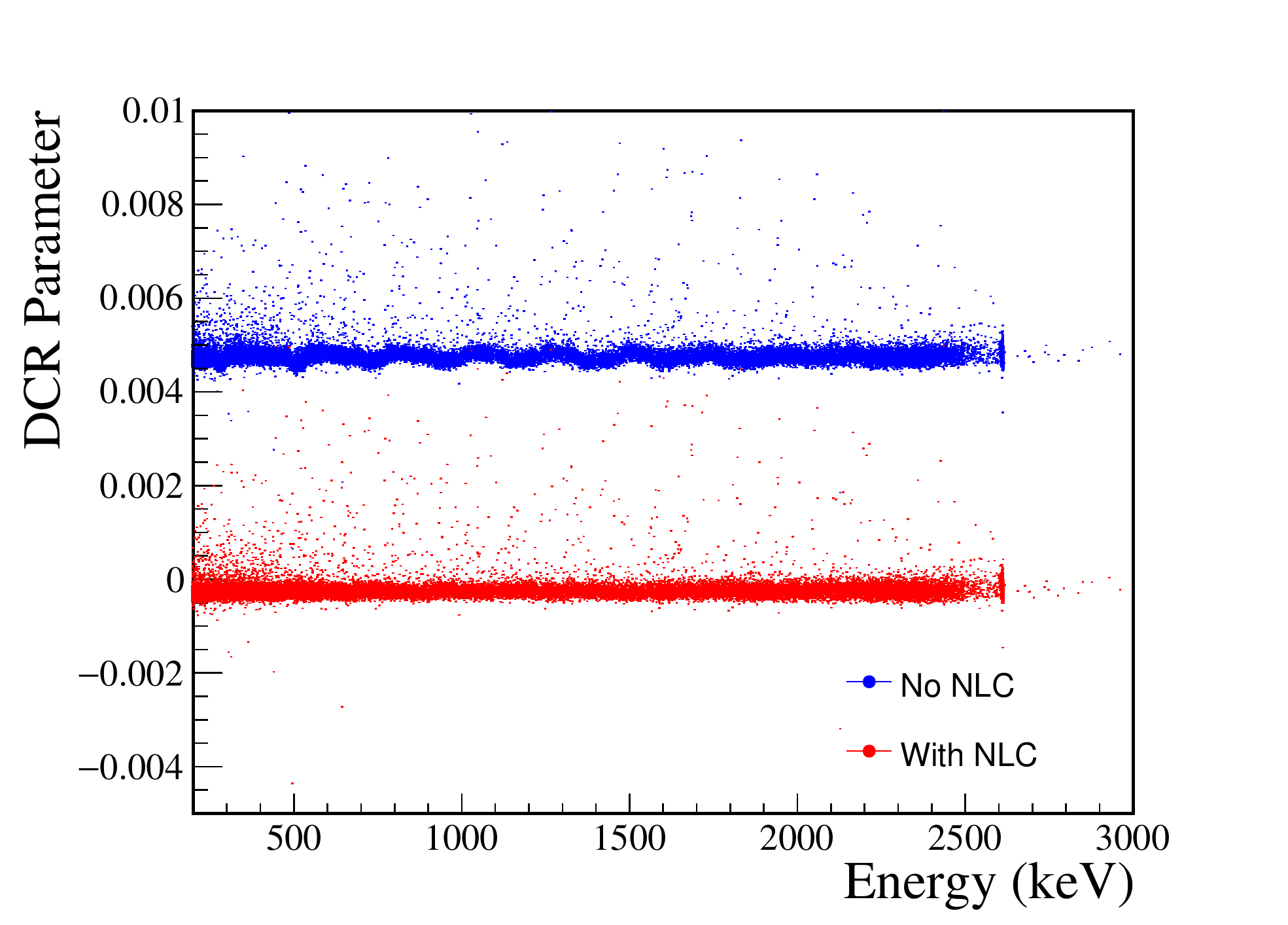}
\caption{ADC nonlinearities observed in the delayed charge recovery parameter for discriminating alphas, measured with calibration data in one detector. The blue dots show how the ADC nonlinearity injects a visible wiggle in the DCR parameter as a function of energy, while the red dots demonstrate how this variation is removed by the nonlinearity correction.  An offset of the blue dots is added for better visualization. }
\label{fig:NL_DCR}
\end{figure}

Finally, Fig.~\ref{fig:NL_DCR} shows the impact of ADC nonlinearities and their correction in the DCR parameter used to reject events from alpha particles striking the passivated surface of the detectors~\cite{Gruszko:2017}. Alphas incident on the passivated surface of the HPGe detectors exhibit significant charge trapping that is partially recovered at delayed times relative to the fast rise of their pulses, which is measured as a significantly positive DCR. A cut is chosen to select events consistent with DCR = 0, with the acceptance for non-alpha events tuned to be near 99\%. The ADC nonlinearity injects a visible wiggle in the DCR parameter as a function of energy. Correcting for the nonlinearity significantly improves the DCR resolution and therefore improves discrimination against alpha-incident events while maintaining high efficiency. The nonlinearity correction also reduces the energy-dependence of the DCR cut signal acceptance. The detailed impact of the nonlinearity correction on the DCR acceptance is obvious but not straightforward to estimate because of other factors such as the DCR dependence on the  drift time.

\section{Energy Uncertainty due to ADC Nonlinearity}

Ultimately ADC nonlinearities cannot be completely eliminated, and their residual systematic effects must be characterized. In this final section, we describe the quantification of the contribution of residual ADC nonlinearities to the systematic uncertainty in event energy estimation in the \MJD. Systematic uncertainties in other pulse shape parameters can be estimated using similar techniques.

The impact of nonlinearities on energy estimation can be quantified in terms of their effect on the detector response function. In the \MJD\ this is particularly relevant, because the physics is extracted via the search for a peak at a known energy (the $^{76}$Ge double-beta decay spectral endpoint, 2039~keV), with the shape of the response function. In the \MJD, as for many detectors, the peak shape is predominantly Gaussian, so that it is characterized essentially by just two parameters: the mean and RMS width. ADC nonlinearities modify both of these parameters. We compute the systematic uncertainties in these two parameters assuming an analysis in which events from all crystals are combined into a single distribution; biases for a crystal-by-crystal analysis can be computed using identical techniques.

The primary impact of ADC nonlinearities is an energy-dependent shift in the mean of the response function due directly to the value of the (residual) nonlinearity at any given energy. We refer to this as the ``local'' energy nonlinearity. We quantify the size of the local energy nonlinearities using the energy differences estimated between high and low gain channels ($\Delta_{LH} (E)$), shown in Fig.~\ref{fig:NL_HGLG}. We divide the energy spectrum into fine energy bins so that the nonlinearity can be assumed to be constant over the bin. We then compute the average energy difference between low and high gains in each bin for each detector. Next, to accommodate our choice of an all-crystal analysis, we compute the average ($\overline{\Delta}_{LH}(E)$) and the standard deviation ($\sigma_{\Delta}(E)$) of these energy differences over all of the detectors in the array. The former biases the mean of the detector response, and the latter contributes to its width.

\begin{figure}[htbp]
\centering
\includegraphics[width=0.45\textwidth]{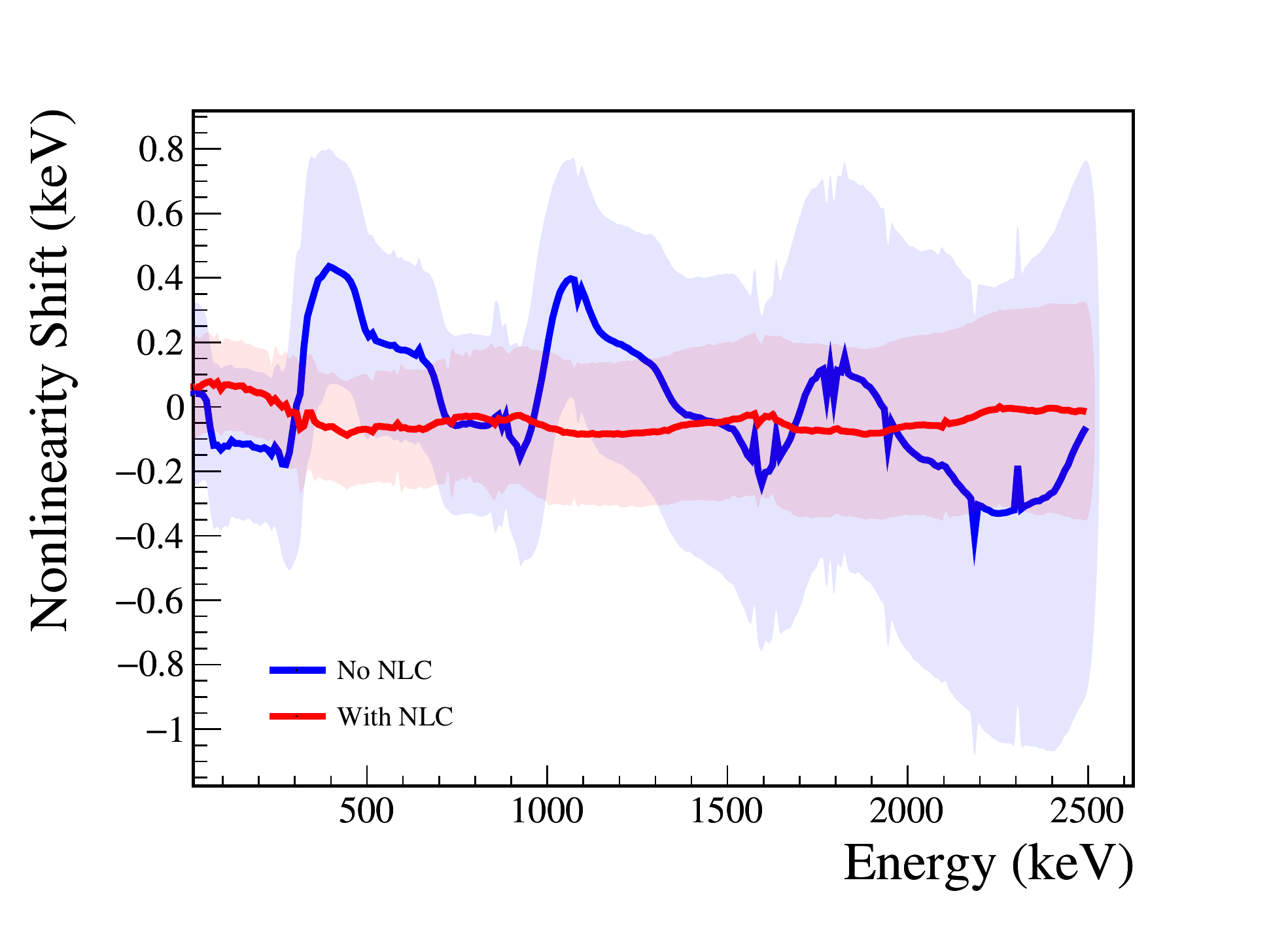}
\caption{Local energy systematic bias (solid line) and additional variance (shaded region) due to residual nonlinearities before and after the nonlinearity correction for all operating high gain channels using high-statistics calibration data. }
\label{fig:NL_corr}
\end{figure}

The energy variations in the $\Delta_{LH}(E)$ are observed to be roughly constant in scale across the entire energy spectrum. Since high gain channels in the \MJD\ have $\sim$3 times the gain of low gain channels, we assume that the relative residual nonlinearity at a given energy differ by a factor of 3 on average. Depending on the relative sign at each energy, these nonlinearities may add to or subtract from each other to give the observed $\overline{\Delta}_{LH}(E)$. We thus expect the residual nonlinearities to lie within the ranges $\overline{\Delta}_{LH} (E)/\left(1 \pm 3\right) $ for high gain channels, and $\overline{\Delta}_{LH}(E)/\left(1 \pm \frac{1}{3}\right) $ for low gain channels. We conservatively estimate the corresponding contributions to the systematic uncertainty in the detector response function mean to be $\frac{1}{2} \overline{\Delta}_{LH}(E)$ for high gain channels, and $\sqrt{1 + \left(\frac{1}{3}\right)^2} \overline{\Delta}_{LH}(E)$ for low gain channels.  The solid lines in Fig.~\ref{fig:NL_corr} shows the trend for $\frac{1}{2} \overline{\Delta}_{LH}(E)$ as a function of energy for all operating high gain channels using high-statistics calibration data with and without the nonlinearity correction applied. The energy uncertainty is highly suppressed by the nonlinearity correction and is below 0.1~keV over the full calibration energy range. 

In addition to shifting the mean, when events from multiple detectors are combined into a single spectrum, detector-to-detector differences in the local energy nonlinearities contribute to additional width in the combined, array-wide response function.  Unfortunately, we cannot disentangle any difference in variability between the low gain and high gain nonlinearities from the data we have, so we conservatively take the full variability as an estimate of the resulting increase in the energy width for either gain. We thus inflate the width parameter of the energy response function by adding it in quadrature with $\sigma_{\Delta}(E)$ directly. The trend of $\sigma_\Delta(E)$ before and after nonlinearity correction is plotted as the shaded regions in Fig.~\ref{fig:NL_corr}. The width is substantially reduced by the correction, especially at higher energies, and is at the level of $\sim$0.2~keV over the entire energy range. The uncertainty contribution to the energy response function width due to the uncertainty in $\sigma_\Delta(E)$ ($<$ 0.01 keV) is negligible compared to the Fano width ($\sim 0.2$ keV).

In the \MJD\ and any similarly calibrated detector, nonlinearities also have a more global effect on the energy scale: the nonlinearities at the calibration points lead to non-statistical fit residuals (see Fig.~\ref{fig:NL_calibration}), representing a possible pull in the entire calibration. Without a full model of the detailed energy dependence of the residual nonlinearities, this global energy bias can be bracketed with an additional systematic uncertainty in the mean of the detector response function. This contribution can be incorporated directly into the calibration fits by simply inflating the uncertainties in the calibration fit parameters by the square-root of the reduced $\chi^2$ of the fit. This corresponds to adding an additional global variance at each calibration point that accounts for their non-statistical scatter about the best-fit curve. As can be seen from the scatter of the data points in Fig.~\ref{fig:NL_calibration}, in the \MJD\ this contribution to the energy uncertainty is on the order of 0.05~keV.

ADC nonlinearity may also depend on the environment including temperature, power supply and aging changes within the ADCs. However, the  measurement was done in the underground laboratory where the environment is reasonably stable for the rare decay experiment. Gain stability and nonlinearity stability studies also show very small fluctuations for the experiment over the last few years. Since ADC nonlinearities are presumed to be a static property of the ADCs and since no evidence is observed for significant time variation of $\Delta_{LH} (E)$, possible time-dependent changes are ignored.

\section{Conclusion}
Nonlinearity is a well-known issue for fast ADCs. In this paper, we have shown this effect is observable in various aspects of \MJD\ data, including the energy difference between the low and high gains, the energy calibration residuals,  and the delayed charge recovery pulse shape discrimination parameter. We have also demonstrated that the nonlinearity exhibits a non-trivial hysteresis and yet can be measured with inexpensive signal generators and corrected with simple, efficient algorithms. After the nonlinearity correction, the energy deviation from linearity is less than 0.1~keV, and the additional contribution to the energy width is about 0.2~keV. This correction was required to achieve the record energy resolution of the \MJD's neutrinoless double-beta decay search.

\section{Acknowledgment}

We thank our hosts and colleagues at the Sanford Underground Research Facility for their support. 

N. Abgrall, C.M.~Campbell, Y-D.~Chan, H.~L.~Crawford, A.~Drobizhev, J.~MyslikA and A.~W.~P.~Poon are with the Nuclear Science Division, Lawrence Berkeley National Laboratory, Berkeley, CA 94720, USA.

J.~M.~Allmond, F.~E.~Bertrand, V.~E.~Guiseppe, D.~C.~Radford, R.~L.~Varner and C.-H.~Yu are with the Oak Ridge National Laboratory, Oak Ridge, TN 37830, USA.

I.~J.~Arnquist, E.~W.~Hoppe and  R.~T.~Kouzes are with Pacific Northwest National Laboratory, Richland, WA 99354, USA.

F.~T.~Avignone~III is with the Department of Physics and Astronomy, University of South Carolina, Columbia, SC 29208, USA and the Oak Ridge National Laboratory, Oak Ridge, TN 37830, USA.

A.~S.~Barabash is with National Research Center ``Kurchatov Institute'' Institute for Theoretical and Experimental Physics, Moscow, 117218 Russia.

C.~J.~Barton, J.~M.~L\'opez-Casta\~no, T.~K.~Oli and W.~Xu are with the Department of Physics, University of South Dakota, Vermillion, SD 57069, USA.

B.~Bos, T.~S.~Caldwell, M.~L.~Clark, J.~Gruszko, I.~S.~Guinn, R.~J.~Hegedus, C.~R.~Haufe, R.~Henning, D.~Hervas~Aguilar, E.~L.~Martin, G.~Othman, J.~Rager and A.~L.~Reine are with the Department of Physics and Astronomy, University of North Carolina, Chapel Hill, NC 27599, USA and Triangle Universities Nuclear Laboratory, Durham, NC 27708, USA.

M.~Busch is with the Department of Physics, Duke University, Durham, NC 27708, USA and Triangle Universities Nuclear Laboratory, Durham, NC 27708, USA.

M.~Buuck was with the Center for Experimental Nuclear Physics and Astrophysics, and Department of Physics, University of Washington, Seattle, WA 98195, USA. He is now with the SLAC National Accelerator Laboratory, Menlo Park, CA 94025, USA.

C.~D.~Christofferson is with South Dakota School of Mines and Technology, Rapid City, SD 57701, USA.

P.-H.~Chu, S.~R.~Elliott, I.~Kim, R.~Massarczyk, S.~J.~Meijer, K.~Rielage, M.~J.~Stortini, B.~R.~White are with Los Alamos National Laboratory, Los Alamos, NM 87545, USA.

C.~Cuesta was with the Center for Experimental Nuclear Physics and Astrophysics, and Department of Physics, University of Washington, Seattle, WA 98195, USA. She is now with Centro de Investigaciones Energ\'{e}ticas, Medioambientales y Tecnol\'{o}gicas, CIEMAT 28040, Madrid, Spain.

J.~A.~Detwiler, A.~Hostiuc, N.~W.~Ruof and C.~Wiseman are with the Center for Experimental Nuclear Physics and Astrophysics, and Department of Physics, University of Washington, Seattle, WA 98195, USA.

D.~W.~Edwins and D.~Tedeschi are with the Department of Physics and Astronomy, University of South Carolina, Columbia, SC 29208, USA.

Yu.~Efremenko is with the Department of Physics and Astronomy, University of Tennessee, Knoxville, TN 37916, USA and the Oak Ridge National Laboratory, Oak Ridge, TN 37830, USA.

H.~Ejiri is with the Research Center for Nuclear Physics, Osaka University, Ibaraki, Osaka 567-0047, Japan.

T.~Gilliss was with the Department of Physics and Astronomy, University of North Carolina, Chapel Hill, NC 27599, USA and Triangle Universities Nuclear Laboratory, Durham, NC 27708, USA. He is now with the Applied Physics Laboratory, Johns Hopkins University, Laurel, MD 20723, USA.

G.~K.~Giovanetti is with the Physics Department, Williams College, Williamstown, MA 01267, USA.

M.~P.~Green is with the Department of Physics, North Carolina State University, Raleigh, NC 27695, USA, Triangle Universities Nuclear Laboratory, Durham, NC 27708, USA and Oak Ridge National Laboratory, Oak Ridge, TN 37830, USA.

M.~F.~Kidd is with the Department of Physics, Tennessee Tech University, Cookeville, TN 38505, USA.

A.~M.~Lopez is with the Department of Physics and Astronomy, University of Tennessee, Knoxville, TN 37916, USA.

R.~D.~Martin is with the Department of Physics, Engineering Physics and Astronomy, Queen's University, Kingston, ON K7L 3N6, Canada.

S.~Mertens is with Max-Planck-Institut f\"{u}r Physik, M\"{u}nchen, 80805 Germany and Physik Department and Excellence Cluster Universe, Technische Universit\"{a}t, M\"{u}nchen, 85748, Germany.

W. Pettus was with the  Center  for  Experimental  Nuclear Physics  and  Astrophysics,  and  Department  of  Physics,  University  of  Washington,  Seattle,  WA  98195,  USA. He is now with the Department of Physics, Indiana University, Bloomington IN 47405 and IU Center for Explorationof Energy and Matter, Bloomington IN 47408, USA.

J.~Rager was with the Department of  Physics  and  Astronomy,  University of North  Carolina, Chapel  Hill, NC 27599, USA  and Triangle  Universities  Nuclear Laboratory, Durham, NC 27708, USA. He is now with  Applied Research Associates, Raleigh NC 27615, USA.

S.~Vasilyev is with Joint Institute for Nuclear Research, Dzhelepov Laboratory of Nuclear Problems, Dubna, 141980 Russia.

J.~F.~Wilkerson is with the Department of Physics and Astronomy, University of North Carolina, Chapel Hill, NC 27599, USA, Triangle Universities Nuclear Laboratory, Durham, NC 27708, USA and Oak Ridge National Laboratory, Oak Ridge, TN 37830, USA.

B.~X.~Zhu was with Los Alamos National Laboratory, Los Alamos, NM 87545, USA. He is now with Jet Propulsion Laboratory, California Institute of Technology, Pasadena, CA 91109, USA.

B.~Shanks was with the Oak Ridge National Laboratory, Oak Ridge, TN 37830 USA.

\bibliography{main}
\bibliographystyle{IEEEtran} 
\end{document}